\documentclass[a4paper,english,aps,prl,twocolumn,floatfix,showpacs,amsfonts,amssymb,superscriptaddress]{revtex4} 
\usepackage{graphicx}
\usepackage{amssymb}
\usepackage{amsmath}

\begin{document}

\title{Density functional theory with adaptive pair density}

\author{J. Lorenzana}
\affiliation{ISC-CNR and Department of Physics, Sapienza 
University of Rome, Piazzale Aldo Moro 2, I-00185, Rome, Italy}
\author{Z.-J. Ying}
\affiliation{ISC-CNR and Department of Physics, Sapienza 
University of Rome, Piazzale Aldo Moro 2, I-00185, Rome, Italy}
\author{V. Brosco}
\affiliation{ISC-CNR and Department of Physics, Sapienza 
University of Rome,
 Piazzale Aldo Moro 2, I-00185, Rome, Italy}

\date{\today}

\begin{abstract}
We propose a density functional to find the ground state energy and
density of interacting particles,  where both the density and the pair density
can adjust in the presence of an inhomogeneous 
potential. 
As a proof of principle we formulate an {\sl a priori} exact functional   
for the inhomogeneous Hubbard model.
The functional has the same form as the Gutzwiller approximation but with
an unknown kinetic energy reduction factor.   
An approximation to the functional based on the exact solution of the uniform
problem leads to a substantial improvement over the local density
approximation. 
\end{abstract}


\maketitle

Most of our theoretical understanding of condensed matter and complex
molecules stems from density functional theory (DFT)
computations \cite{Kohn1999Nobel}.  Practical implementations rely on
the local density approximation (LDA) or  its gradient generalizations
which often provide accurate results 
at the modest cost of a Hartree like computation \cite{fio03}. These methods however fail
in systems at or close to the Mott insulating
regime \cite{Imada1998Metalinsulator}, when the tunneling matrix
element of electrons becomes small compared with the typical electron-electron 
repulsion energies.  This can be seen already at the level of
an H$_2$ molecule which is stretched to produce two separated H
atoms. LDA performs reasonably well at the equilibrium distance, when
electrons have substantial tunneling among the atoms, but  
fails in the molecular analog of the Mott regime, when each   
electron is localized on one H atom \cite{bae97,coh08}. 

Generalizing to the local spin density approximation improves the
energy at the cost of an artificial breaking of the
symmetry.\cite{gun76} While this can 
be formally justified,\cite{Perdew1995Escaping} in practice it leads to unwanted
features. For example in the case of a strongly correlated metal 
the artificial breaking of symmetry will give rise to a Fermi surface 
with the wrong Luttinger volume.\cite{Luttinger1960Fermi}

The breakdown of unpolarized LDA in the H molecule 
can be traced back to the poor treatment of each H atom
separately.\cite{bae97,coh08} 
The density close to the center of the H atom  is 0.32 a.u. 
LDA essentially assumes that this portion
of the system behaves as a uniform electron gas with the same density 
(Wigner-Seitz radius parameter  $r_s\sim 0.91$). However the two
systems have radically different pair distribution functions which
leads to different electron-electron interaction 
energies (zero for the H atom). LDA thus introduces a spurious interaction
of the electron with itself, the so-called self-interaction error \cite{per81}. A way to mitigate this problem
would be to have a functional theory which depends  both 
on the density and the pair density (DPDFT)
 so that it can discriminate between situations with the same density but
different pair densities. 

The idea to involve the pair density in electronic structure
computations is older than DFT
itself \cite{Coulson1960Present,Coleman1963Structure}.
More recently various works explored the possibility to define an energy functional based on the pair-density alone
 \cite{Ziesche1994Pair,Furche2004Towards,lev01,Schuch2009Computational}  
or  explore the possibility to adjust the 
spatially averaged pair density\cite{GoriGiorgi2006Systemadapted} but
face serious problems on finding physically acceptable pair
densities.\cite{Schuch2009Computational}
Our proposal differs in that we still keep the density as the basic
variable but we use partial information on the pair density as an
auxiliary variable which gives the functional more sensitivity to
correlation. In this respect our functional is similar to the 
proposal of Ref.~\cite{Perdew1995Escaping} with the difference that
 it does not need an artificial breaking of symmetry. 

To show the feasibility and the usefulness of the formalism in strongly
correlated systems we present a DPDFT  for the
one-dimensional Hubbard model { \sl i.e.} electrons on a lattice with
strong local interaction. The functional is inspired on the Gutzwiller
approximation (GA) \cite{gut63,vol84,geb90}
in the same way as Kohn-Sham approach is inspired on
the Hartree approximation.\cite{koh65} 
It is thus in principle exact as Kohn-Sham theory 
represents  the ``exactification'' of the Hartree
approximation\cite{Kohn1999Nobel}.
We develop an approximation to the unknown functional 
which becomes exact for uniform systems and  
involves local or semilocal quantities like LDA and its gradient
generalizations, but which is more accurate and goes beyond LDA in the
sense that it becomes highly non-local when expressed as a standard
functional of the density alone.  

We consider a one-dimensional inhomogeneous 
Hubbard model $H=H_t+H_U+H_v$ with 
$$H=-t\sum_{x\sigma}(c_{x\sigma }^{\dag}c_{x+1\sigma
}+{\rm H.c.})+\sum_x U_x n_{x\uparrow }n_{x\downarrow }
+\sum_{x\sigma}v_xn_{x\sigma},$$
here $c_{x\sigma}$  and $c^{\dag}_{x\sigma}$
indicate the annihilation and creation operator of electrons with spin
$\sigma$ on site $x$ and  $n_{x\sigma
}\equiv c^{\dag}_{x\sigma}c_{x\sigma}$.
$t$ is the nearest-neighbor hopping amplitude while 
$U_x$ and $v_x$ denote respectively the Hubbard interaction
energy and the external potential on site $x$. For reasons to become
clear below we allow the interaction energy to be site dependent. 

The Hohenberg and Kohn
theorem \cite{hoh64,Schonhammer1995Densityfunctional} guarantees that
there exists a functional, 
 $E_v[\{\rho_{x }\}]=F[\{U_x,\rho _{x}\}]+\sum_{x}v_x\rho_x$,   
which, when minimized with respect to the density provides the exact ground state energy.
$F[\{U_x,\rho _{x}\}]$  is a ``universal'' functional independent of
$v_x$, but depending on the $U_x$'s, which  represents the contribution
of $H_t+H_U$ to the energy of a system with density $\rho_{x}$.
A formal DPDFT for this model can be obtained by performing the Legendre
transform,\cite{aye09} 
\begin{equation}
  \label{eq:tdn}
  T[\{d_x,\rho_{x}\}]=\max_{\{U_x\}} \left( F[\{U_x,\rho_{x}\}]-\sum_x U_x d_x\right)
\end{equation}
where $d_x=\langle n_{x\uparrow }n_{x\downarrow } \rangle$ is the
on-site pair density or double occupancy. Identifying the last term in
the brackets with the interaction energy we arrive at the conclusion that
$T[\{d_x,\rho_{x}\}]$ is the interacting kinetic energy of the model
with the specified pair density and density distributions. This is a
universal functional which does not depend on the specific form of
$U_x$ nor $v_x$. Below we develop an approximation for this functional.

The ground state energy of the system is obtained by minimizing the
functional,
\begin{equation}
  \label{eq:newf}
 E_{Uv}[\{d_x,\rho_{x }\}]= T[\{d_x,\rho_{x}\}]+\sum_{x}U_xd_x
 +\sum_{x}v_x\rho_x    
\end{equation}
with respect to $d_x$ and $\rho_{x }$. In the new scheme $H_t$ has become
the fixed part of the 
Hamiltonian while both $H_U$ and  $H_v$ are considered as problem dependent. 
 We will show below that even in the
conventional case in which $U_x$ is homogeneous the new functional
is quite convenient. 

 Eq.~(\ref{eq:tdn}) yields, 
$$\frac{\partial F[\{U_x,\rho_{x}\}]}{\partial U_{x'}}=d_{x'}.$$ 
Inversion of this expression determines the set of  $U_x$ a system must have
to have the given $d_x$ and $\rho _{x}$. Since $H_v$ and  $H_U$ are
determined by   
$\rho_x$ and $d_x$, the wave-function and all physical quantities are
functionals of $\rho_x$ and $d_x$.

The interacting kinetic energy can be written as
$$T=-t\sum_{x}(\rho_{x,x+1}+\rho_{x+1,x})$$
where  $\rho_{x,x'}$ denotes the one body density matrix,   $\rho_{x,x'}\equiv\sum_\sigma\langle  c_{x\sigma }^{\dag}c_{x'\sigma}
\rangle$ and is also a functional of $\rho_x$ and $d_x$.

To proceed we search for a non-interacting system  
satisfying the, to be determined, Kohn-Sham like equations, 
\begin{equation}
\label{eq:kohnsham}
-\sum_{\delta=\pm 1}t^{\rm eff}_{x,x+\delta}
\varphi_{x+\delta\sigma}^{(\nu)}+ v^{\rm eff}_x \varphi _{x\sigma} ^{(\nu
)}=\varepsilon_\nu \varphi_{x\sigma}^{(\nu )},
\end{equation}
and which has the same density of the interacting system.
Defining the non-interacting one-body density matrix as the sum restricted to the occupied Kohn-Sham orbitals,
\begin{equation}
\label{eq:rho0}
\rho_{x,x^{\prime }}^0=\sum_{\sigma\nu\in occ. }\varphi_\sigma ^{*(\nu
)}(x)\varphi _\sigma^{(\nu )}(x'), 
\end{equation}
and the density as  $\rho_{x}^0\equiv\rho_{x,x}^0$, we require thus that,
$  \rho_{x}=\rho_{x}^0$.

In addition we introduce the kinetic energy 
reduction factors which are also functionals of $d$ and $\rho$ and satisfy 
\begin{equation}
  \label{eq:q}
\rho_{x,x\pm 1}=q_{x,x\pm 1}[\{d_{x'},\rho_{x' }\}]\rho_{x,x\pm 1}^0.
 \end{equation}

With these definitions the kinetic energy functional can be written in
terms of the non interacting density matrix as follows
\begin{equation}
  \label{eq:kine}
T[\{d_x,\rho_{x}^0\}]=-t \sum_{x',\delta=\pm1}
q_{x',x'+\delta}[\{d_x,\rho_{x}^0\}]\rho_{x',x'+\delta}^0.
\end{equation}
Similarly to the exchange correlation potential in the
standard Kohn-Sham approach, $q$ 
encodes all the complication of the many-body problem in an exact way.

Minimizing Eqs.~(\ref{eq:newf}) with respect to the orbitals and using Eq. \eqref{eq:kine} we arrive at
\begin{eqnarray}
t^{\rm eff}_{x,x\pm 1}&=&t q_{x,x\pm 1}[\{d_{x'},\rho_{x'}^0\}],\\ 
v^{\rm eff}_x&=&v_x-t \sum_{x',\delta}\frac{ \partial
  q_{x',x'+\delta}}{\partial \rho_x} \rho^0_{x',x'+\delta} .  
  \label{eq:teffveff}
\end{eqnarray}

Eqs.~(\ref{eq:kohnsham})-(\ref{eq:teffveff}) define the kinetic energy
reduction functional
$q_{x,x+\delta}[\{d_{x'},\rho_{x'}\}]$ and the associated Kohn-Sham
system.

For practical computations one needs to introduce 
approximations. In analogy with the  Gutzwiller
approximation\cite{gut63,geb90}
we assume the kinetic energy reduction functional  
factorizes in terms of functions of the local densities,
$q_{x,y}=z(\rho_x,d_x)z(\rho_y,d_y)$. In the following we refer to this approximation as ``factorized approximation'' (FA).
 In the spirit of LDA,\cite{koh65,Lima2003Density} we
approximate the functional dependence of $z(\rho_x,d_x)$ by requesting that the
functional yields the exact energy in the case of a uniform
system which we obtain in one-dimension by solving
numerically the exact Bethe-ansatz integral equations\cite{lie68}
(in higher dimension a numerical solution can be used). 
Thus $z(d,\rho)$ is calculated from the condition
$z^2(d,\rho)T_0(\rho )=T^{BA}(d,\rho ),$ where
  $T_0(\rho )$ denotes the kinetic energy of the non-interacting
uniform system. 
$T^{BA}(d,\rho )$ is obtained as a function of the double occupancy
$d$, from the exact Bethe-ansatz energy
$E(U,\rho)$ 
by the Legendre transform Eq.~(\ref{eq:tdn}) for a system with uniform
$U$ and density $\rho$.  It is also useful to compute the function
$U^p(\rho,d)$ which yields the Hubbard $U$ a uniform
system with density $\rho=\rho_x$
must have, to yield the double occupancy $d_x$ of the non-uniform
system at the given site. 
We indicate as  $d(\rho,U^p)$ the inverse of $U^p(\rho,d)$.

The functional Eq.~(\ref{eq:newf}) has to be minimized with respect to 
the local double occupancies $d_x$, and the 
Kohn-Sham orbitals leading to Eq.~(\ref{eq:kohnsham}). 
We can use the function $d(\rho,U^p)$ to eliminate 
$d_x$ in favor of a site dependent effective interaction, termed the
``pseudointeraction'', $U_x^p$ with respect to which the minimization is
actually done [$d_x\rightarrow d(\rho_x,U_x^p) $]. 
This change of variables  allows us to avoid the problem of minimizing
with respect to a constrained variable.
 In the case of a homogeneous external potential $v_x=V$, the minimum
is attained when $U_x^p=U$, and one recovers the exact Bethe ansatz
result.

DPDFT is a variant of DFT. Indeed once a functional for $T$ is known
we can define a functional  of the density alone by minimizing over $d_x$, i.e. $F[\{U,\rho_{x}\}]= \min_{\{d_x\}} 
\left( T[\{d_x,\rho_{x}\}]+U \sum_x  d_x\right)$. Interestingly, as we know from previous
works\cite{sei01}, this procedure leads to a highly non-local  
functional of the density starting from a 
nearly local functional of  variables $\rho_x$, $d_x$. 

\begin{figure}[!t]
\begin{center}
\includegraphics[width=9 cm]{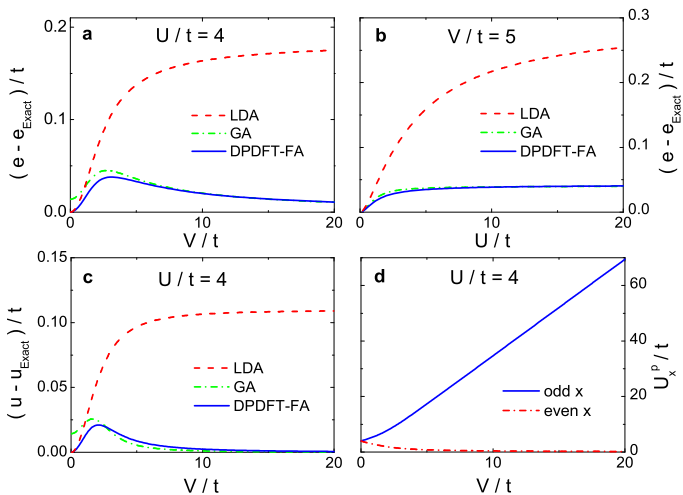}
\end{center}
\vskip -.6 cm
\caption{(a) Error on the 
energy per site, $e$, for a periodic Hubbard chain of length  $L=16$ with a
 binary potential and $N=6$ electrons as a function
of the potential strength $V$. 
Notice that as for the LDA,  DPDFT-FA   becomes asymptotically exact
for a negligible external potential 
but as the GA approximation gives a small error in the case of a
strong inhomogeneous potential where LDA fails. (b) Error  
as a function of the interaction
strength.  (c) Error in the interaction energy per
site $u= U\sum_x d_x/L$ as a function of $V$. (d) The
pseudointeraction (defined in the text) at odd and even sites
 as a function of the potential strength. 
} 
\label{fig1IHM}
\end{figure}

In order to test the functional we solved the 
DPDFT-FA equations for a Hubbard chain with a binary potential  
$v_x=(-1)^x V$ (known as ionic Hubbard
model\cite{Nagaosa1986Theory,Egami1993Lattice})
and compare it with standard LDA derived from the exact Bethe ansatz
solution\cite{Lima2003Density} and with exact results obtained by Lanczos diagonalization
with the ALPS package\cite{alps}.

Fig.~\ref{fig1IHM}  shows that DPDFT-FA performs much better than LDA
when the system becomes highly inhomogeneous (a) and strongly
interacting (b). The double occupancy in LDA is independent of the
environment which  
leads to a poor approximate interaction energy for strongly
inhomogeneous systems (c). In DPDFT-FA the double occupancy is
allowed to adapt, so that localized electrons, for large values of the
potential, tend to have a small double occupancy and a small
interaction energy (c), explaining the better
performance of DPDFT-FA respect to LDA. 
 The enhanced  pseudointeraction on the more charged sites [odd $x$ in
  (d)] leads to a reduced double occupancy and explains the small interaction
energy.

The behavior of DPDFT-FA is qualitatively similar to the GA except at small $V$
where the GA is obviously not exact.  
The errors in other quantities like kinetic energy, potential energy
and density (not shown) are also generically 
smaller in DPDFT-FA than in LDA.

These results suggest that the self-interaction (SI) should be strongly 
reduced in  DPDFT-FA with respect to LDA. In order to verify that this is
the case we have solved the problem of one electron with one attractive 
impurity site in a chain.  The potential is given by
$v_x=-V \delta_{x,0}$. This is the lattice 
analogue of the hydrogen atom in the continuum. 
Being a single electron problem, the exact
solution has $d_x=0$ for all $U$ while both LDA and  DPDFT-FA yield a
finite $d_x$. We define the self-interaction error as the spurious
interaction energy of the single electron problem, $E_{SI}=U \sum_x d_x$.

\begin{figure}[b]
\begin{center}
\includegraphics[width=8cm]{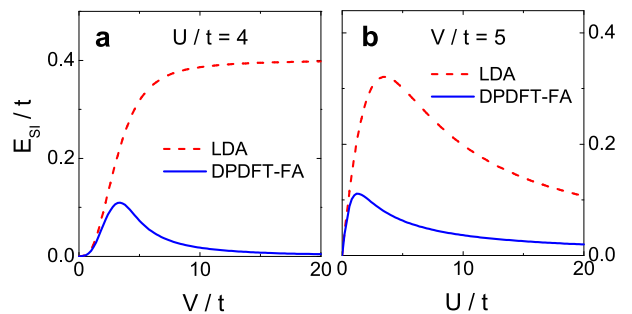}
\end{center}
\vskip -.5 cm
\caption{Self-interaction error for a chain of $L=12$ sites, a 
single electron  and a potential in which one site has strength
$-V$ as a  function of $V$ (a) and $U$ (b). For large
  potential strength the density at the impurity becomes close to
  one. In LDA this fixes the double occupancy to an unphysical value
  while in DPDFT-FA the double occupancy adjusts to small values
  when the electron localizes leading to a small self-interaction.
 } \label{fig-SIC}
\end{figure}

In Fig.~\ref{fig-SIC}(a) we plot the self-interaction error as a
function of the potential strength at the impurity site, with 
interaction parameter $U=4t$. For large $V$ the charge becomes
localized at the impurity site with $\rho_0\sim 1$. 
In LDA the interaction energy corresponds to that of a
nearly half-filled uniform Hubbard model which is clearly a very bad
approximation thus $E_{SI}$ (red dashed line) increases and tends
to saturate at a large value. In DPDFT-FA  (blue solid line)  $E_{SI}$
starts with a slower increase and then decays to very small values. 
In this case the total double occupancy  $E_{SI}/U$ becomes small
showing the adaptability of the pair density to the local environment. 
As shown in Fig.~\ref{fig-SIC}(b), 
the reduction of the self-interaction error is
large for a wide range of the interaction.  

It is interesting to compare DPDFT with traditional Kohn-Sham DFT for the same
model\cite{Lima2003Density}. In the latter 
the difference between the interacting and the non-interacting kinetic
energy is absorbed additively in the exchange-correlation
potential. Here the correction is multiplicative and included in the
kinetic energy reduction factor functional $q$.

Formally the minimization in DFT has to be restricted 
to densities, and here also double occupancies,  that correspond to a physical
wave function (N-representability).  While this may appear as a severe
difficulty\cite{Schuch2009Computational} it is often not a problem in practical
implementations. It has indeed not hampered the development of DFT
methods. In our case the problem is not more severe than in LDA
because each portion of the system is approximated by a uniform system
but with a modified interaction.

We have shown DPDFT at work in the lattice but a similar functional
can be defined in the continuum. Ideally one would like to use the pair
density, $\gamma_{\sigma\sigma'}({\bf r},{\bf r}')=\langle
\psi^\dagger_\sigma({\bf r}) \psi^\dagger_{\sigma'} ({\bf
  r}')\psi_{\sigma'} ({\bf r}') \psi_{\sigma} ({\bf r})
\rangle $, as a variable, with $\psi_{\sigma} ({\bf r})$ the field
operator at point 
${\bf r}$ with spin $\sigma$. 
This, however, would be quite cumbersome as in each point a full function
must be determined. A more practical approach is to impose a
spatial dependent constraint on the pair density and use
such a constraint to parameterize families of physical pair density
functions. For example we can define, 
$$D({\bf r})=\sum_{\sigma\sigma'}\int 
d^3{\bf r}'\gamma_{\sigma\sigma'}({\bf r},{\bf r}')  \theta(a-|{\bf
  r}-{\bf r'}|), $$
 with $\theta$ the Heaviside function and $a$ is an
appropriately chosen cutoff radius. $D$ measures the double occupancy
probability within a sphere of radius $a$. 
 These or other constraints can be implemented\cite{aye09} 
by replacing the physical
 interaction $w({\bf  r},{\bf r'})$ with a fictitious interaction. 
In
 the present example the fictitious interaction would read  
$w({\bf  r},{\bf r'})+ U({\bf r})\theta(a-|{\bf
  r}-{\bf r'}|)  $. As in the lattice, the
corresponding Hohenberg and Kohn functional has  
$U({\bf r})$ as a variable,  $F[n({\bf r}),U({\bf r})]$
which allows to define the functional 
\begin{equation}
  \label{eq:gdn}
\nonumber
  G[n({\bf r}),D({\bf r})]=\max_{U({\bf r})} 
\left( F[n({\bf r}),U({\bf r})]-\int  d^3{\bf r} U({\bf r}) D({\bf r})\right).
\end{equation}
leading to a theory where both $n({\bf r})$ and $D({\bf r})$ are
fundamental variables similar to the lattice but with the difference
that $G$ is not the kinetic energy. Solution of the uniform problem
in the presence of the fictitious interaction with a constant $U({\bf
  r})=U^p$ could serve as a basis for approximate functionals which
converge to the LDA in the uniform case but have an adaptive exchange
correlation hole in non-uniform situations.

We have conceptually shown how a DFT  which uses the pair
density as an auxiliary variable can be introduced and we have developed an 
approximation for the Hubbard model inspired on the GA 
combined with LDA ideas. 
Formally DPDFT can also be defined in the continuum with a local
variable $D({\bf r})$ playing the role of the double occupancy in the
lattice. Approximate functionals based on this approach
should allow more control on the correlations built on the underlying
 wave-function respect to what LDA does, and could help to  
extend the success of DFT methods to strongly correlated systems where 
correlations are substantially different from those of the homogeneous
electron gas.

 We thank A. Filippetti and  V. Fiorentini for
  useful discussions. We are in debt with  P. Gori-Giorgi for
a  critical riding of the manuscript and many valuable suggestions. 
This work was supported by IIT-Seed project NEWDFESCM.


\begin{thebibliography}{10}

\bibitem{Kohn1999Nobel}
W. Kohn, Rev. Mod. Phys. {\bf 71},  1253  (1999).

\bibitem{fio03}
{\em A Primer in Density Functional Theory (Lecture Notes in Physics) (v.
  620)}, 1 ed., edited by C. Fiolhais, F. Nogueira, and M. Marques (Springer,
  Berlin, 2003).

\bibitem{Imada1998Metalinsulator}
M. Imada, A. Fujimori, and Y. Tokura, Rev. Mod. Phys. {\bf 70},  1039
   (1998).

\bibitem{bae97}
E.~J. Baerends and O.~V. Gritsenko, J. Phys. Chem. A {\bf 101},  5383  (1997).

\bibitem{coh08}
A.~J. Cohen, P. Mori-S\'{a}nchez, and W. Yang, Science {\bf 321},  792  (2008).

\bibitem{gun76}
O. Gunnarsson and B.~I. Lundqvist, Phys. Rev. B {\bf 13},  4274  (1976).

\bibitem{Perdew1995Escaping}
J.~P. Perdew, A. Savin, and K. Burke, Phys. Rev. A {\bf 51},  4531
  (1995).

\bibitem{Luttinger1960Fermi}
J.~M. Luttinger, Phys. Rev. {\bf 119},  1153
  (1960).

\bibitem{per81}
J.~P. Perdew and A. Zunger, Phys. Rev. B {\bf 23},  5048  (1981).

\bibitem{Coulson1960Present}
C.~A. Coulson, Rev. Mod. Phys. {\bf 32},  170  (1960).

\bibitem{Coleman1963Structure}
A.~J. Coleman, Rev. Mod. Phys. {\bf 35},  668  (1963).

\bibitem{Ziesche1994Pair}
P. Ziesche, Phys. Lett. A {\bf 195},  213  (1994).

\bibitem{Furche2004Towards}
F. Furche, Phys. Rev. A {\bf 70},  022514  (2004).

\bibitem{lev01}
M. Levy and P. Ziesche,  J. Chem. Phys. {\bf 115},  9110
  (2001).
  
 \bibitem{Schuch2009Computational}
N. Schuch and F. Verstraete, Nature Phys. {\bf 5},  732  (2009).


\bibitem{GoriGiorgi2006Systemadapted}
P. Gori-Giorgi and A. Savin, Phil. Mag. {\bf 86},  2643  (2006).


\bibitem{gut63}
M.~C. Gutzwiller, Phys. Rev. Lett. {\bf 10},  159  (1963).

\bibitem{vol84}
D. Vollhardt, Rev. Mod. Phys. {\bf 56},  99  (1984).

\bibitem{geb90}
F. Gebhard, Phys. Rev. B {\bf 41},  9452  (1990).

\bibitem{koh65}
W. Kohn and L.~J. Sham, Phys. Rev.  {\bf 140}, A1133  (1965).

\bibitem{Schonhammer1995Densityfunctional}
K. Sch\"{o}nhammer, O. Gunnarsson, and R.~M. Noack, Phys. Rev. B {\bf 52},
   2504  (1995).

\bibitem{hoh64}
P. Hohenberg and W. Kohn, Phys. Rev.  {\bf 136},
  B864  (1964).

\bibitem{aye09}
P.~W. Ayers and P. Fuentealba, Phys. Rev. A {\bf 80},  032510  (2009).

\bibitem{Lima2003Density}
N.~A. Lima, M.~F. Silva, L.~N. Oliveira, and K. Capelle, Phys. Rev.
  Lett. {\bf 90},  146402  (2003). Differently from this reference to
  define the LDA we
  did not used an approximate formula for the  energy but
  worked with the exact Bethe ansatz expression. 

\bibitem{lie68}
E.~H. Lieb and F.~Y. Wu, Phys. Rev. Lett. {\bf 20},  1445  (1968).

\bibitem{sei01}
G. Seibold and J. Lorenzana, Phys. Rev. Lett. {\bf 86},  2605  (2001).

\bibitem{Nagaosa1986Theory}
N. Nagaosa and J.-I. Takimoto, J. Phys. Soc. Japan {\bf 55},
  2735  (1986).

\bibitem{Egami1993Lattice}
T. Egami, S. Ishihara, and M. Tachiki, Science {\bf 261},  1307  (1993).

\bibitem{alps} B. Bauer {\em et al.} J. Stat. Mech. P05001 (2011); F. Albuquerque {\sl et al.}, J. Mag. and Mag. Mat. {\bf 310}, 1187 (2007).

\end{thebibliography}

\end{document}